# Effect of receiver orientation on resource allocation in optical wireless systems

Osama Zwaid Alsulami[1], Khulood D. Alazwary[1], Sarah O. M. Saeed[1], Sanaa Hamid Mohamed[1],
T. E. H. El-Gorashi[1], Mohammed T. Alresheedi[2] and Jaafar M. H. Elmirghani[1]
[1]School of Electronic and Electrical Engineering, University of Leeds, LS2 9JT, United Kingdom
[2]Department of Electrical Engineering, King Saud University, Riyadh, Kingdom of Saudi Arabia
ml15ozma@leeds.ac.uk, elkal@Leeds.ac.uk, elsoms@leeds.ac.uk, elshm@leeds.ac.uk,
t.e.h.elgorashi@leeds.ac.uk, malresheedi@ksu.edu.sa, j.m.h.elmirghani@leeds.ac.uk

**ABSTRACT**
Optical wireless communication (OWC) systems have been the subject of a significant amount of interest as they can be used in sixth generation (6G) wireless communication to provide high data rates and support multiple users simultaneously. This paper investigates the impact of receiver orientation on resource allocation in optical wireless systems, using a wavelength division multiple access (WDMA) scheme. Three different systems that have different receiver orientations are examined in this work. Each of these systems considers 8 simultaneous users in two scenarios. WDMA is utilised to support multiple users and is based on four wavelengths offered by Red, Yellow, Green and Blue (RYGB) LDs for each AP. An angle diversity receiver (ADR) is used in each system with different orientations. The optimised resource allocations in terms of wavelengths and access point (AP) is obtained by using a mixed-integer linear programming (MILP) model. The channel bandwidth and SINR are determined in the two scenarios in all systems. The results show that a change in the orientation of the receiver can affect the level of channel bandwidth and SINR. However, SINRs in both scenarios for all users are above the threshold (15.6 dB). The SINR obtained can support t data rate of 5.7 Gbps in both scenarios in all systems.

## 1. INTRODUCTION

With upcoming networks, such as the Internet of Things (IOT), the amount of Internet-connected devices is increasing exponentially [1]. However, the radio frequency (RF) spectrum which is currently used in many wireless applications is becoming insufficient due to its inherent limitations, such as spectrum scarcity, and the demand for higher data rates. Optical wireless communications (OWC) is a technology that can potentially support and meet the growing demand for high data rates and has been the subject of interest of researchers, being proposed as a part of sixth generation communication (6G) systems. Recently, the Visible Light Communication (VLC) system, as a type of OWC system, has been the subject of a lot of research due to its high capacity and high level of security compared to radio frequrncy (RF) wireless systems [2], [3]. Furthermore, many demonstration works have shown that VLC systems can have capacities of up to 25 Gbps in indoor environments [4]–[12].

In addition, to improve the communication link, different techniques such as beam adaptation can be applied together with other adaptation techniques such as power, angle and delay adaptation [3], [13]–[19]. Another improvement in the link can be achieved by using diversity techniques, such as angle diversity receivers [20], [21] to reduce interference. Uplink OWC systems were studied in [22], [23], however, energy efficiency must be given more consideration [24]. A multi-carrier code division multiple access (MC-CDMA) scheme was studied to support multi-user OWC [13], [25]. However, interference between users is one of the impairments in the VLC system when considering multiple users; and it affects the performance of the system. Therefore, many techniques have been proposed to reduce interference by using various orthogonal resources, such as time, wavelength, and frequency [12], [26]–[30]. Wavelength division multiple access (WDMA) is one solution that can offer multi-user access in VLC systems while reducing interference.

In this paper, the impact of receiver orientation on wavelengths and access point (AP) allocation is studied. Laser Diodes (LDs) consisting of four wavelengths – Red, Yellow, Green, and Blue (RYGB) – are used in the optical transmitter. They offer white colour for the purpose of indoor illumination as well as high bandwidth modulation for the purpose of communication [31]. Thus, the WDMA scheme is utilised to offer multiple access. Unlike previous studies that have assumed a fixed ADR orientation [9], [10], this work uses three different orientations (three systems) based on ADRs. The optimum allocation of wavelengths and access points is identified by maximizing all users' SINRs, using a Mixed Integer Linear Programme (MILP).

The remainder of this paper is organised as follows: the system configuration, including the room configuration, transmitter configuration, receiver configuration and ADR rotation scenarios are introduced in Section 2. The simulation results are discussed in Section 3, while the conclusions are presented in Section 4.

## 2. SYSTEM CONFIGURATION

In this work, an empty room was considered with no doors or windows (See Figure 1). The room includes 8 light units (access points) that provide illumination and communication. In the simulation, the optical indoor channel was modelled using a ray tracing algorithm similar to [32]-[36]. Only light of sight, first and second order reflections were included in the simulation as the higher order reflections have no significant impact on the received power [33]. Therefore, room surfaces (ceiling, walls and floor) are divided into small identical areas (elements) that act as secondary emitters which reflect the light rays in the form of a Lambertian pattern, as shown in [5]. These elements play a significant role in the resolution of the results. When the elements' size in each surface is very small, higher temporal resolution results are obtained, but this may lead to increase in the computation time needed for the simulation. All communication links operate above the communication floor (CF), as shown in Figure 1.

An angle diversity receiver (ADR) with four branches similar to [9] was used, as shown in Figure 2. Each branch in the ADR has a photodetector with narrow Field of View (FOV) to collect signals and reduce interference. Moreover, the orientation of each detector can be selected to cover a different sector of the room by using two angles: Azimuth ($Az$) and Elevation ($El$). In this work, three different orientations of the ADR were examined. Table 1 illustrates the simulation parameters of the room, transmitter and receiver.

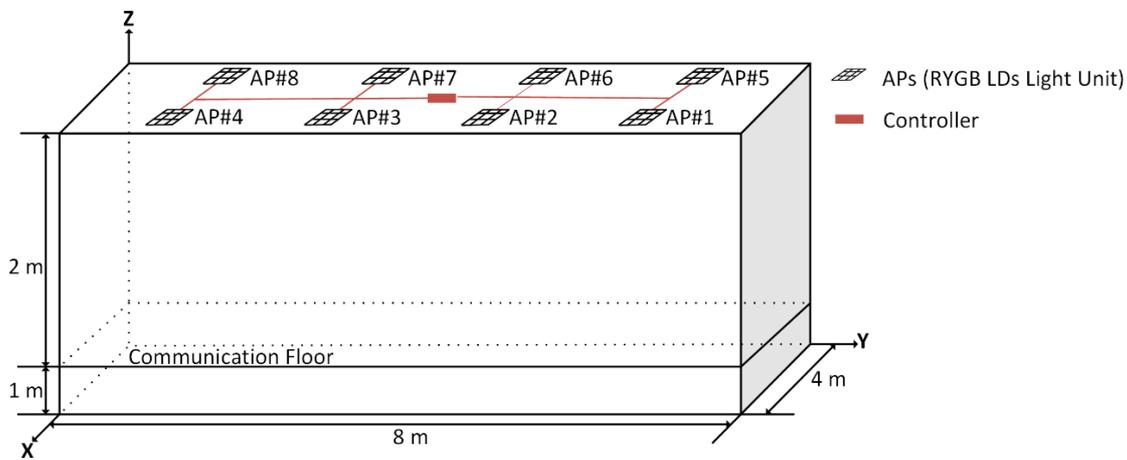

Figure 1: Room Configuration.

**Table 1.** System Parameters

| Parameters | Configurations | |
|---|---|---|
| Walls and ceiling reflection coefficient | 0.8 [33] | |
| Floor reflection coefficient | 0.3 [33] | |
| Number of reflections | 1 | 2 |
| Area of reflection element | 5 cm × 5 cm | 20 cm × 20 cm |
| Order of Lambertian pattern, walls, floor and ceiling | 1 [33] | |
| Semi-angle of reflection element at half power | 60º | |
| Number of RYGB LDs per unit | 12 | |
| Transmitted optical power of Red LD | 0.8 W | |
| Transmitted optical power of Yellow LD | 0.5 W | |
| Transmitted optical power of Green LD | 0.3 W | |
| Transmitted optical power of Blue LD | 0.3 W | |
| Total transmitted power of RYGB LD | 1.9 W | |
| **Room** | | |
| Width × Length × Height (x, y, z) | 4 m × 8 m × 3 m | |
| Number of transmitters' units | 8 | |

| Transmitters locations (x, y, z) | (1 m, 1 m, 3 m), (1 m, 3 m, 3 m), (1 m, 5 m, 3 m), (1 m, 7 m, 3 m), (3 m, 1 m, 3 m), (3 m, 3 m, 3 m), (3 m, 5 m, 3 m) and (3 m, 7 m, 3 m) | | | |
|---|---|---|---|---|
| **Receiver** | | | | |
| Responsivity Red | 0.4 A/W | | | |
| Responsivity Yellow | 0.35 A/W | | | |
| Responsivity Green | 0.3 A/W | | | |
| Responsivity Blue | 0.2 A/W | | | |
| Number of photodetectors | 4 | | | |
| Area of the photodetector | 20 mm$^2$ | | | |
| Receiver noise current spectral density | 4.47 pA/√Hz [5] | | | |
| Receiver bandwidth | 4 GHz | | | |
| **System 1** | | | | |
| Photodetector | 1 | 2 | 3 | 4 |
| Azimuth angles | 0° | 90° | 180° | 270° |
| Elevation angles | 60° | 60° | 60° | 60° |
| Field of view (FOV) of each detector | 25° | | | |
| **System 2** | | | | |
| Photodetector | 1 | 2 | 3 | 4 |
| Azimuth angles | 30° | 120° | 210° | 300° |
| Elevation angles | 60° | 60° | 60° | 60° |
| Field of view (FOV) of each detector | 25° | | | |
| **System 3** | | | | |
| Photodetector | 1 | 2 | 3 | 4 |
| Azimuth angles | 60° | 150° | 240° | 330° |
| Elevation angles | 60° | 60° | 60° | 60° |
| Field of view (FOV) of each detector | 25° | | | |

## 3. SIMULATION SETUP AND RESULTS

In this work, laser diodes (LDs) consisting of four wavelengths – Red, Yellow, Green and Blue (RYGB) – are used in each AP for providing illumination and communication. The four wavelengths are mixed, providing white light and enabling it to be used in the indoor environment, as reported in [31]. The multiple access scheme used in this work is wavelength division multiple access (WDMA). Three different system configurations have been evaluated in this work each configuration represents a different receiver orientation. In each system, two different 8-user scenarios were considered. In the first scenario, each four users were located under one AP. However, in the second scenario, all users were distributed over the room and each user was located under one AP. A mixed-integer linear programming (MILP) model was used to optimise the resource allocation by maximising the sum of SINRs of all users [12], [26]. The users' locations in both scenarios and their optimised assignment for the three systems are illustrated in Tables 2 and 3. Moreover, it should be noted that the users' locations are known to the controller that is placed on the ceiling of the room as shown in Figures 1 and 2; [12], [26]. Figure 2 shows an example to clarify the WDMA resource allocation using MILP in a VLC system scenario that consists of three APs, three users and two wavelengths (Red and Blue). The solid lines in Figure 2 refer to the desired modulated link to a user that is assigned to specific AP and wavelength, while dashed lines indicate an interference modulated link that uses the same wavelength but a different AP. Dotted lines specify the unmodulated wavelength (background noise) from another AP. As shown in the example in Figure 2, User 1 is assigned to AP1 using the blue wavelength and has been affected only by the background noise, while Users 2 and 3 are assigned to AP2 and AP3 respectively with the same wavelength (Red) and are affected by interference and background noise.

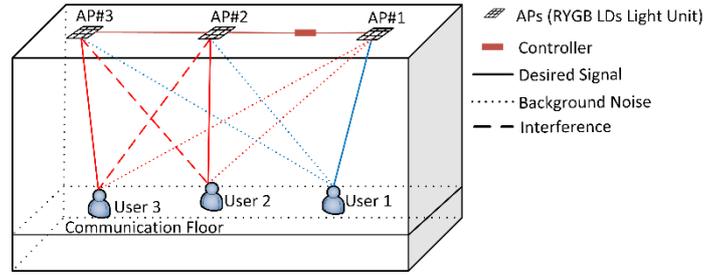

Figure 2: WDMA example.

**Table 2.** Scenario 1 with the optimised resource allocation

| User | Location (x, y, z) | System 1 | | | System 2 | | | System 3 | | |
|---|---|---|---|---|---|---|---|---|---|---|
| | | AP | Branch | wavelength | AP | Branch | wavelength | AP | Branch | wavelength |
| 1 | (0.5,6.5,1) | 4 | 1 | Red | 4 | 1 | Red | 3 | 4 | Red |
| 2 | (0.5,7.5,1) | 4 | 3 | Yellow | 4 | 4 | Yellow | 4 | 4 | Yellow |
| 3 | (1.5,6.5,1) | 3 | 3 | Red | 3 | 3 | Red | 8 | 1 | Red |
| 4 | (1.5,7.5,1) | 8 | 4 | Red | 8 | 4 | Red | 4 | 3 | Red |
| 5 | (2.5,0.5,1) | 1 | 2 | Red | 1 | 2 | Red | 5 | 1 | Red |
| 6 | (2.5,1.5,1) | 6 | 1 | Red | 6 | 1 | Red | 1 | 3 | Red |
| 7 | (3.5,0.5,1) | 5 | 2 | Yellow | 5 | 2 | Yellow | 5 | 2 | Yellow |
| 8 | (3.5,1.5,1) | 5 | 3 | Red | 5 | 3 | Red | 6 | 2 | Red |

**Table 3.** Scenario 2 with the optimised resource allocation

| User | Location (x, y, z) | System 1 | | | System 2 | | | System 3 | | |
|---|---|---|---|---|---|---|---|---|---|---|
| | | AP | Branch | wavelength | AP | Branch | wavelength | AP | Branch | wavelength |
| 1 | (0.5,1.5,1) | 1 | 4 | Red | 1 | 4 | Red | 1 | 4 | Yellow |
| 2 | (0.5,5.5,1) | 3 | 3 | Yellow | 3 | 4 | Yellow | 3 | 4 | Yellow |
| 3 | (0.5,6.5,1) | 4 | 1 | Red | 4 | 1 | Red | 4 | 1 | Yellow |
| 4 | (1.5,3.5,1) | 2 | 3 | Red | 2 | 3 | Red | 2 | 3 | Red |
| 5 | (2.5,1.5,1) | 5 | 4 | Red | 5 | 4 | Red | 5 | 4 | Red |
| 6 | (2.5,6.5,1) | 8 | 1 | Red | 8 | 1 | Red | 8 | 1 | Red |
| 7 | (3.5,3.5,1) | 6 | 3 | Yellow | 6 | 3 | Yellow | 6 | 3 | Yellow |
| 8 | (3.5,5.5,1) | 7 | 4 | Red | 7 | 3 | Red | 7 | 3 | Red |

The optimisation of the resource allocation (APs and wavelengths) to each user for both scenarios in the three is obtained by using the MILP model. Following the optimum allocation of resources, the channel bandwidth and the SINR at a fixed data rate for each user were determined as illustrated in Figures 3, 4.

In Figure 3, the optical channel bandwidth is shown for each user in both scenarios in the three systems. The optical channel bandwidth varies between the three systems. System 3 in Scenario 2 has the best channel bandwidth compared to Systems 1 and 2. Also, the orientation of the ADR can affect the channel bandwidth due to the change in coverage area of each detector and the assignment of the AP which may increase the inter-symbol interference (ISI), leading to a decrease in the channel bandwidth as shown in Tables 2 and 3.

Figure 4 shows the SINR for each user in both scenarios across all systems. The SINR is evaluated at a fixed data rate of 5.7 Gbps. This data rate was chosen because of the limited channel bandwidth of some users' locations, which assumed a minimum value of 4 GHz. Systems 1 and 2 show almost an identical SINR compared to system 3. They provide higher SINR for users 1 and 8, while system 3 offers high SINR for users 4 and 5 in scenario 1. In scenario 2, systems 1 and 2 have better SINR for some users compared to system 3. However, all users in both

Scenarios for all systems provide a high SINR which is above the threshold (15.6 dB) needed for $P_e=10^{-9}$ for On Off Keying (OOK).

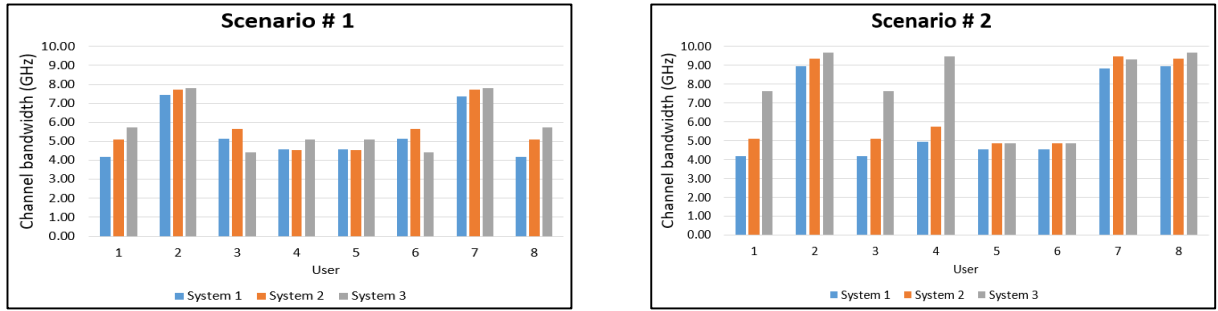

Figure 3: Channel bandwidth.

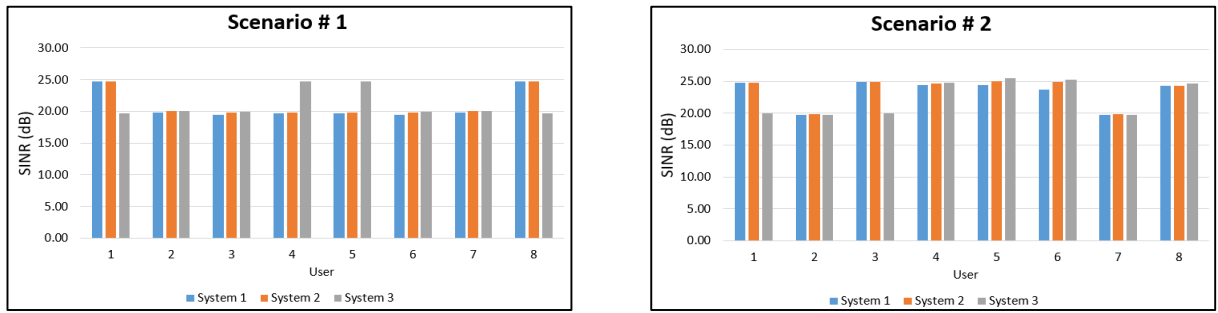

Figure 4: SINR.

## 4. CONCLUSIONS

The impact of receiver orientation on resource allocation in optical wireless systems was investigated in this paper using a WDMA scheme. Three different systems were evaluated in this work examining different receiver orientations. Each system considered 8 users in two scenarios. A WDMA scheme was used to support multiple users based on four wavelengths offered from RYGB LDs for each AP in the VLC system. An ADR was used in each system with different orientations. The optimisation of resource allocation in terms of wavelengths and APs was achieved using a mixed-integer linear programming (MILP) model. Each user channel bandwidth and SINR were calculated in the two scenarios for all systems. The results show that the channel bandwidth and SINR change when the receiver orientation changes and this affects the optimum allocation of wavelengths and APs to each receiver. All users in both Scenarios for all systems offer a high SINR above the threshold (15.6 dB). The SINR was evaluated at a data rate of 5.7 Gbps in both scenarios in all systems.

## ACKNOWLEDGEMENTS

The authors would like to acknowledge funding from the Engineering and Physical Sciences Research Council (EPSRC) INTERNET (EP/H040536/1), STAR (EP/K016873/1) and TOWS (EP/S016570/1) projects. The authors extend their appreciation to the deanship of Scientific Research under the International Scientific Partnership Program ISPP at King Saud University, Kingdom of Saudi Arabia for funding this research work through ISPP#0093. OZA would like to thank Umm Al-Qura University in the Kingdom of Saudi Arabia for funding his PhD scholarship, KDA would like to thank King Abdulaziz University in the Kingdom of Saudi Arabia for funding her PhD scholarship, SOMS would like to thank the University of Leeds and the Higher Education Ministry in Sudan for funding her PhD scholarship. SHM would like to thank EPSRC for providing her Doctoral Training Award scholarship. All data are provided in full in the results section of this paper.## REFERENCES

1. Cisco Mobile, *Cisco Visual Networking Index: Global Mobile Data Traffic Forecast Update, 2016-2021 White Paper*. 2017.
2. Z. Ghassemlooy, W. Popoola, and S. Rajbhandari, *Optical wireless communications: system and channel modelling with Matlab®*. 2012.
3. F. E. Alsaadi, M. A. Alhartomi, and J. M. H. Elmirghani, "Fast and efficient adaptation algorithms for multi-gigabit